\documentclass[preprintnumbers,amsmath,amssymb,floatfix,11pt,prd,twocolumn,superscriptaddress,nofootinbib]{revtex4}
\usepackage{hyperref}
\usepackage{footnote}
\usepackage{amssymb}
\usepackage{amsmath}
\usepackage{graphicx}
\usepackage{dcolumn}
\usepackage{bm}
\usepackage{color}
\usepackage{xcolor}
\usepackage{float}
\usepackage{lipsum}
\allowdisplaybreaks
\usepackage{subfig}
\usepackage{comment}
\usepackage{multirow}
\usepackage[figurename=FIGURE,tablename=TABLE]{caption}
\usepackage{threeparttable}
\usepackage{hyperref}
\hypersetup{
    colorlinks=true, %set true if you want colored links
    linktoc=all,     %set to all if you want both sections and subsections linked
    linkcolor=magenta,
    citecolor=red
}

\newcommand{\ba}{\begin{eqnarray}}
\newcommand{\ea}{\end{eqnarray}}

\begin{document}

\title{ACT Constraints on Marginally Deformed Starobinsky Inflation}

\author{Jureeporn Yuennan}
\email{jureeporn\_yue@nstru.ac.th}
\affiliation{Faculty of Science and Technology, Nakhon Si Thammarat Rajabhat University, Nakhon Si Thammarat, 80280, Thailand}

\author{Farruh Atamurotov}
\email{atamurotov@yahoo.com}
\affiliation{Urgench State University, Kh. Alimdjan str. 14, Urgench 220100, Uzbekistan}

\author{Phongpichit Channuie}
\email{phongpichit.ch@mail.wu.ac.th}
\affiliation{School of Science, Walailak University, Nakhon Si Thammarat, 80160, Thailand}
\affiliation{College of Graduate Studies, Walailak University, Nakhon Si Thammarat, 80160, Thailand}

\date{\today}% It is always \today, today,

\begin{abstract}
We investigate the inflationary phenomenology of a marginally deformed Starobinsky model, motivated by quantum corrections to the $R^{2}$ term, in light of the latest cosmological observations. In this framework, the inflationary potential acquires a small deformation parameter, $\gamma$, which shifts predictions away from the exact Starobinsky limit. Using the slow-roll formalism, we derive analytic expressions for the spectral index $n_{s}$ and tensor-to-scalar ratio $r$ and confront them with constraints from Planck, ACT, and DESI data. Our analysis shows that nonzero values of $\gamma$ raise both $n_{s}$ and $r$, thereby alleviating the $\gtrsim 2\sigma$ tension between the Starobinsky $R^{2}$ scenario and the ACT+DESI (P-ACT-LB) measurements, which favor $n_{s} \simeq 0.9743 \pm 0.0034$. For $N \sim 60$ $e$-foldings, the model consistently reproduces the observed amplitude of primordial perturbations while predicting tensor contributions within current observational bounds. We also demonstrate that the deformation softens the otherwise severe fine-tuning of the quartic self-coupling in minimally coupled inflation. The parameter range $\gamma \sim \mathcal{O}(10^{-3})$–$\mathcal{O}(10^{-2})$ emerges as phenomenologically viable, providing a natural extension of Starobinsky inflation compatible with present data. We conclude that marginally deformed $R^{2}$ inflation remains a compelling and testable candidate for the primordial dynamics of the Universe, with future CMB and gravitational-wave observations expected to further probe its parameter space.

\end{abstract}

%\keywords{Traversable Wormholes}

\maketitle

%\tableofcontents
\newpage
%%%%%%%%%%%%%%%%%%%%%%%%%%%%
\section{Introduction}
%%%%%%%%%%%%%%%%%%%%%%%%%%%%
Recently, the Atacama Cosmology Telescope (ACT) data~\cite{ACT:2025fju,ACT:2025tim} 
combined with the DESI data~\cite{DESI:2024uvr,DESI:2024mwx} made the scientific community to 
reconsider the benchmark primordial theory of our Universe, that is inflation, 
since the ACT data indicated that the scalar spectral index of the primordial 
curvature perturbations is in at least $2\sigma$ discordance with the Planck 
data~\cite{Planck:2018jri}. Inflation has become a cornerstone of modern cosmology, offering a compelling 
resolution to the flatness, horizon, and monopole problems of the standard 
Big Bang scenario. Moreover, it naturally explains the generation of primordial 
perturbations, which served as the seeds of large-scale structure and are 
observed today as anisotropies in the cosmic microwave background (CMB)~\cite{Starobinsky:1980te,Sato:1981qmu,Guth:1980zm,Linde:1981mu,Albrecht:1982wi}. 
These fluctuations are usually characterized by two key observables: the scalar 
spectral index, $n_s$, describing the scale dependence of scalar modes, and the 
tensor-to-scalar ratio, $r$, measuring the amplitude of primordial gravitational 
waves relative to scalar perturbations.  

For a chosen inflationary potential, both quantities can typically be expressed 
in terms of the number of $e$-foldings $N$ between horizon exit and the end of 
inflation. This framework allows precise theoretical predictions to be compared 
against observational data. A particularly notable outcome is the universal 
relation $n_s = 1 - \tfrac{2}{N}$, which is realized across a wide range of 
models. These include $\alpha$-attractor scenarios~\cite{Kallosh:2013tua,Kallosh:2013hoa,Kallosh:2013maa,Kallosh:2013yoa,Kallosh:2014rga,Kallosh:2015lwa,Roest:2015qya,Linde:2016uec,Terada:2016nqg,Ueno:2016dim,Odintsov:2016vzz,Akrami:2017cir,Dimopoulos:2017zvq}, 
the $R^2$ model of Starobinsky inflation~\cite{Starobinsky:1980te}, and Higgs 
inflation with large nonminimal coupling to gravity~\cite{Kaiser:1994vs,Bezrukov:2007ep,Bezrukov:2008ej}. 
Similar predictions also arise in models with composite inflaton fields~\cite{Karwan:2013iph,Channuie:2012bv,Bezrukov:2011mv,Channuie:2011rq}, 
as reviewed in~\cite{Channuie:2014ysa,Samart:2022pza}. For the benchmark value 
$N = 60$, this universal form gives $n_s \approx 0.9667$, which aligns well with 
the {\it Planck} 2018 result $n_s = 0.9649 \pm 0.0042$~\cite{Planck:2018jri}.  

However, more recent ACT measurements~\cite{ACT:2025fju,ACT:2025tim}, especially 
when combined with other probes, point toward a higher scalar spectral index than 
inferred by Planck alone. A joint analysis of ACT and Planck (P-ACT) yields 
$n_s = 0.9709 \pm 0.0038$, while including CMB lensing and baryon acoustic 
oscillation data from DESI (P-ACT-LB) further increases the estimate to 
$n_s = 0.9743 \pm 0.0034$. These updated constraints put significant pressure on the universal attractor 
class of models, effectively ruling them out at about the $2\sigma$ level and 
raising serious challenges for many inflationary frameworks that predict this 
universal behavior. Ref.~\cite{ACT:2025fju} emphasizes that the P-ACT-LB bounds 
place the Starobinsky $R^2$ model itself under tension at $\gtrsim 2\sigma$. 
This conclusion is both striking and unexpected, in sharp contrast with earlier 
consensus.

There is already a large stream of articles in the cosmology literature that aim to explain the ACT result \cite{Kallosh:2025rni,Gao:2025onc,Liu:2025qca,Yogesh:2025wak,Yi:2025dms,Peng:2025bws,Yin:2025rrs,Byrnes:2025kit,Wolf:2025ecy,Aoki:2025wld,Gao:2025viy,Zahoor:2025nuq,Ferreira:2025lrd,Mohammadi:2025gbu,Choudhury:2025vso,Odintsov:2025wai,Q:2025ycf,Zhu:2025twm,Kouniatalis:2025orn,Hai:2025wvs,Dioguardi:2025vci,Yuennan:2025kde,Oikonomou:2025xms,Oikonomou:2025htz,Odintsov:2025jky,Aoki:2025ywt}. A comprehensive overview of these developments is presented in~\cite{Kallosh:2025ijd}. In the present work, we revisit the quantum-induced marginal deformations of the Starobinsky gravitational action of the form $R^{2(1-\alpha)}$, with $R$ the Ricci scalar and $\alpha$ a positive parameter smaller than one half. This work is organized as follows: In section \ref{II}, we take a short recap of a marginally deformed Starobinsky model, motivated by quantum corrections to the $R^{2}$ term. In
section \ref{III}, we derive the slow-roll parameters and analytic expressions for the inflationary observables including the spectral index $n_{s}$ and tensor-to-scalar ratio $r$. We then in the same section confront them with the recent observational data. Finally, in section \ref{Con}, we summarize our results.

%%%%%%%%%%%%%%%%%%%
\section{Marginally-Deformed Starobinsky Gravity revisited}\label{II}
%%%%%%%%%%%%%%%%%%
An appealing idea is that gravity itself may serve as the driving force behind cosmic inflation.
To investigate this possibility, one must go beyond the standard Einstein–Hilbert (EH) action.
A well-known extension is the Starobinsky model~\cite{Starobinsky:1980te}, in which an $R^{2}$
term is added to the EH action. In this framework, inflation arises naturally from gravity without the need for an additional scalar field.
Remarkably, the model predicts an almost negligible tensor-to-scalar ratio, which is in excellent
agreement with current observational data, such as that from the PLANCK mission~\cite{Ade:2015lrj,Akrami:2018odb}.
Furthermore, logarithmic corrections to the $R^{2}$ term have been suggested in the form
\begin{equation}
\frac{M_{p}^{2}}{2}R + \frac{a}{2}\frac{R^{2}}{1+b\ln(R/\mu^{2})},
\end{equation}
where $R$ denotes the Ricci scalar, $a$ and $b$ are constants, and $\mu$ is a reference energy scale.
Such corrections, motivated by asymptotic safety, have been studied in~\cite{Liu:2018hno}.
From an observational perspective, a potential discovery of primordial tensor modes could strongly constrain the parameters of inflation, expected to lie near the grand unification scale.
In general, the effective gravitational action may be expressed as a Taylor expansion in the Ricci scalar $R$:
\begin{eqnarray}
S &=& \int d^{4}x \sqrt{-g} f(R) \nonumber\\&\equiv&
\int d^{4}x \sqrt{-g}\big(a_{0} + a_{1}R + a_{2}R^{2} + \cdots \big).
\label{eq:action-fR}
\end{eqnarray}
Here $a_{0}$ plays the role of a cosmological constant and must remain small, while $a_{1}$ can be set to unity, as in standard general relativity.
For the Starobinsky model, $a_{2}=1/(6M^{2})$, with $M$ a mass parameter (see~\cite{Chatrabhuti:2015mws} for cosmological implications).
The omitted terms can include contributions from the Weyl tensor $C^{2}$ and the Euler density $E$.
As emphasized in~\cite{Codello:2014sua}, the $E$ term is a total derivative and thus irrelevant, while the Weyl contributions are suppressed in perturbative quantization around flat spacetime.
Since higher powers of $R$, $C^{2}$, and $E$ are Planck-suppressed, they can usually be neglected.
Nonetheless, marginal deformations of~\eqref{eq:action-fR}, realized through logarithmic corrections, have been analyzed in~\cite{Codello:2014sua}.
This leads to a compact Jordan-frame action of the form
\begin{equation}
S_{J} = \int d^{4}x \sqrt{-g}
\bigg[-\frac{M_{p}^{2}}{2}R + h M_{p}^{4\alpha} R^{2(1-\alpha)}\bigg],
\label{eq:Jordan}
\end{equation}
where $h$ is dimensionless and $\alpha$ is a real parameter constrained by $2|\alpha|<1$.
Further discussions of the parameter $\alpha$ can be found in the context of gravity’s rainbow~\cite{Channuie:2019kus}.
To simplify the above form, one can introduce an auxiliary field $y$, rewriting the action as
\begin{equation}
S_{J} = \int d^{4}x \sqrt{-g}\big[f(y) + f'(y)(R-y)\big],
\end{equation}
with
\begin{equation}
f(R) = -\frac{1}{2}M_{p}^{2}R + hM_{p}^{4\alpha} R^{2(1-\alpha)},
\end{equation}
and $f'(y)=df(y)/dy$.
The field equation for $y$ gives $R=y$, provided $f''(y)\neq0$.
A connection to scalar-tensor theories can be established by defining the conformal mode
$\psi = -f'(y)$ and $V(\psi) = -y(\psi)\psi - f(y(\psi))$ and introducing a real scalar $\varphi$ of mass-dimension one through \cite{Codello:2014sua}
\begin{equation}
2\psi - M_{p}^{2} = \xi \varphi^{2}.
\end{equation}
This leads to the alternative Jordan-frame action
\begin{equation}
S_{J} = \int d^{4}x \sqrt{-g}\left[-\frac{M_{p}^{2}+\xi\varphi^{2}}{2}R + V(\varphi)\right],
\label{eq:Jordan2}
\end{equation}
where
\begin{equation}\label{VV}
V(\varphi) = \lambda \varphi^{4}\left(\frac{\varphi}{M_{p}}\right)^{4\gamma},
\qquad \alpha = \frac{\gamma}{1+2\gamma},
\end{equation}
and
\begin{equation}
h^{1+2\gamma} =\Bigg(\frac{\xi}{4}\frac{1+2\gamma}{1+\gamma}\Bigg)^{2(1+\gamma)}\frac{1}{\lambda(1+2\gamma)}.
\label{eq:h-gamma}
\end{equation}
In Eq.~\eqref{eq:Jordan2}, the scalar $\varphi$ lacks a canonical kinetic term.
This can be generated by applying the conformal transformation
\begin{equation}
\tilde{g}_{\mu\nu} = \Omega^{2}(\varphi) g_{\mu\nu}, \qquad
\Omega^{2} = 1 + \frac{\xi\varphi^{2}}{M_{p}^{2}},
\label{eq:conformal}
\end{equation}
which yields the Einstein-frame action
\begin{eqnarray}
S_{E} = \int d^{4}x \sqrt{-g}\bigg[-\frac{M_{p}^{2}}{2}R+
\frac{1}{2}g^{\mu\nu}\partial_{\mu}\chi\partial_{\nu}\chi - U(\chi)\bigg],
\end{eqnarray}
with potential
\begin{equation}
U(\chi) = \Omega^{-4}V(\varphi(\chi)).
\end{equation}
The canonically normalized field $\chi$ is related to $\varphi$ through
\begin{equation}
\frac{1}{2}\left(\frac{d\chi}{d\varphi}\right)^{2} =
\frac{M_{p}^{2}\big(\sigma M_{p}^{2} + (\sigma+3\xi)\xi\varphi^{2}\big)}
{(M_{p}^{2}+\xi\varphi^{2})^{2}}.
\label{eq:chi-phi}
\end{equation}
By setting $\sigma=0$, one recovers the standard mapping between $f(R)$ gravity and its scalar-tensor equivalent. For large values of the non-minimal coupling $\xi$, it is not possible to diﬀerentiate between the
two values of $\sigma=0,\,1$. For large field values $\varphi \gg M_{p}/\sqrt{\xi}$, the relation simplifies to
\begin{equation}
\chi \simeq \kappa M_{p}\ln\left(\frac{\sqrt{\xi}\varphi}{M_{p}}\right)\quad{\rm with}\quad \kappa=\sqrt{\frac{2}{\xi}+6}\,,
\label{eq:chi-large}
\end{equation}
implies that
\begin{equation}
\varphi\rightarrow \frac{M_{p} }{\sqrt{\xi }}\exp\big[\chi/(\kappa M_{p})\big]
\end{equation}
Substituting~\eqref{eq:chi-large} into~\eqref{VV}, the Einstein-frame potential becomes
\begin{equation}
U(\chi) \simeq \frac{\lambda M_{p}^{4}}{\xi^{2}}\left(1+e^{-\frac{2 \chi }{\kappa  M_{p}}}\right)^{-2}\left(\frac{e^{\frac{\chi }{\kappa  M_{p}}}}{\sqrt{\xi }}\right)^{4 \gamma }.
\label{eq:U-chi}
\end{equation}
In the limit $\gamma=0$, one recovers the original Starobinsky potential~\cite{Starobinsky:1980te}. The investigation of inflation in the Einstein frame is quite direct. By applying the standard slow-roll formalism, we evaluate the slow-roll parameters in the large-field regime, using the redefined field $\chi$ and its corresponding potential $U(\chi)$. 

However, it is also convenient to express them in terms of the Jordan frame field $\varphi$ by reinserting (\ref{eq:chi-large}):
\begin{eqnarray}
\varepsilon&=&\frac{M^{2}_{p}}{2}\left(\frac{U'(\chi)}{U(\chi)}\right)^{2}\nonumber\\&=&\frac{2 \left(-2 \gamma +\tanh \left(\frac{\chi }{\kappa M_{p}}\right)-1\right)^2}{\kappa ^2}\nonumber\\&\simeq&\frac{8 M_{p}^4}{\kappa ^2 \xi^2 \varphi ^4}+\frac{16 \gamma M_{p}^2}{\kappa^2 \xi  \varphi ^2}+\frac{8 \gamma ^2}{\kappa ^2}+{\cal O}(\gamma^{3})\\\eta&=& M^{2}_{p}\left(\frac{U''(\chi)}{U(\chi)}\right)\nonumber\\&=&\frac{8}{\kappa^2}\bigg(2 \gamma ^2+\frac{4 \gamma -1}{e^{\frac{2 \chi }{\kappa M_{p}}}+1}+\frac{3}{\left(e^{\frac{2 \chi }{\kappa M_{p}}}+1\right)^2}\bigg)\nonumber\\&\simeq&\frac{8 \left(2 M_{p}^4-M_{p}^2 \xi  \varphi ^2\right)}{\kappa^2 \xi^2 \varphi ^4}+\frac{16 \gamma ^2}{\kappa^2}+\frac{32 \gamma M_{p}^2}{\kappa ^2 \xi  \varphi ^2}\nonumber\\&&+{\cal O}(\gamma^{3}).
\label{eq:U-chi}
\end{eqnarray}
Inflation ends when the slow-roll approximation is violated, in the present case
this occurs for $\varepsilon(\varphi_{\rm end})=1$. Thus the field value at the end of inflation is:
\begin{eqnarray}
\varphi_{\rm end}\simeq \bigg(2^{3/4}+\frac{2 \sqrt[4]{2} \gamma}{\kappa }+\frac{3\ 2^{3/4} \gamma ^2}{\kappa ^2}\bigg)\sqrt{\frac{M_{p}^2}{\kappa  \xi }}\,.\label{phie}
\end{eqnarray} 
We take $\xi \gg 1$, since a value around $\xi \sim 10^{4}$ is necessary to reproduce 
the correct amplitude of density perturbations. This behavior is typical of non-minimally 
coupled single-field inflationary models~\cite{Bezrukov:2007ep,Karwan:2013iph,Channuie:2012bv,Bezrukov:2011mv,Channuie:2011rq,Lee:2014spa,Cook:2014dga}. Although smaller values of $\xi$ are possible, they demand an extremely small $\lambda$, as pointed out in~\cite{Hamada:2014iga}. The quantitative relation between $\xi$ and $\lambda$ will be addressed later, see Eq.~(\ref{phie}).

The Cosmic Microwave Background (CMB) modes that we observe today exited the horizon approximately $N=60$ e-folds prior to the end of inflation. The associated inflaton field value at that moment is denoted by $\chi_{*}$ and is expressed as
\begin{eqnarray}
N&=&\frac{1}{M^{2}_{p}}\int^{\chi_{*}}_{\chi_{\rm end}}\frac{U(\chi)}{dU/d\chi}d\chi\nonumber\\&=&\frac{\kappa ^2 \log \left(1+\gamma  e^{\frac{2 \chi }{\kappa M_{p}}}\right)}{8 \gamma }\Bigg|^{\chi_{*}}_{\chi_{\rm end}}\,.
\end{eqnarray}
In terms of the field $\varphi$, we have
\begin{eqnarray}
\varphi_{*}&\simeq& \frac{M_{p}}{\sqrt{\xi }}\sqrt{\frac{e^{\frac{8 \gamma  N}{\kappa ^2}}-1}{\gamma }}\nonumber\\&\simeq& \Bigg(2 \sqrt{2}+\frac{4 \sqrt{2} \gamma  N}{\kappa^2}+\frac{20 \sqrt{2} \gamma^2 N^2}{3 \kappa^4}\Bigg)\sqrt{\frac{N}{\kappa ^2}}\frac{M_{p}}{\sqrt{\xi}}\nonumber\\&&+{\cal O}(\gamma^{3})\,.
\end{eqnarray}
We performed an expansion in $\gamma$ to illustrate how the outcome departs from the standard $\varphi^{4}$-inflation scenario. The correction induced by $\gamma$ clearly shifts inflation toward larger field values. Nevertheless, such an expansion is valid only when $\gamma$ remains very small. Using $N=60,\,\kappa\sim \sqrt{6}$, we have
\begin{eqnarray}
\varphi_{*}&\simeq& \Big(8.94+178.89\gamma +2981.42\gamma^{2}\Big)\frac{M_{p}}{\sqrt{\xi}}\,.
\end{eqnarray}
Notice that the first term solely displays the contribution of $\varphi^{4}$ model. We observe that the corrections to the quantum correction
parameter of the scalar field, parametrised by $\gamma$, tends to increase the field values of inflation.

%%%%%%%%%%%%%%%%%%%
\section{Confrontation with the ACT Data}\label{III}
\begin{figure}
\includegraphics[width=8 cm]{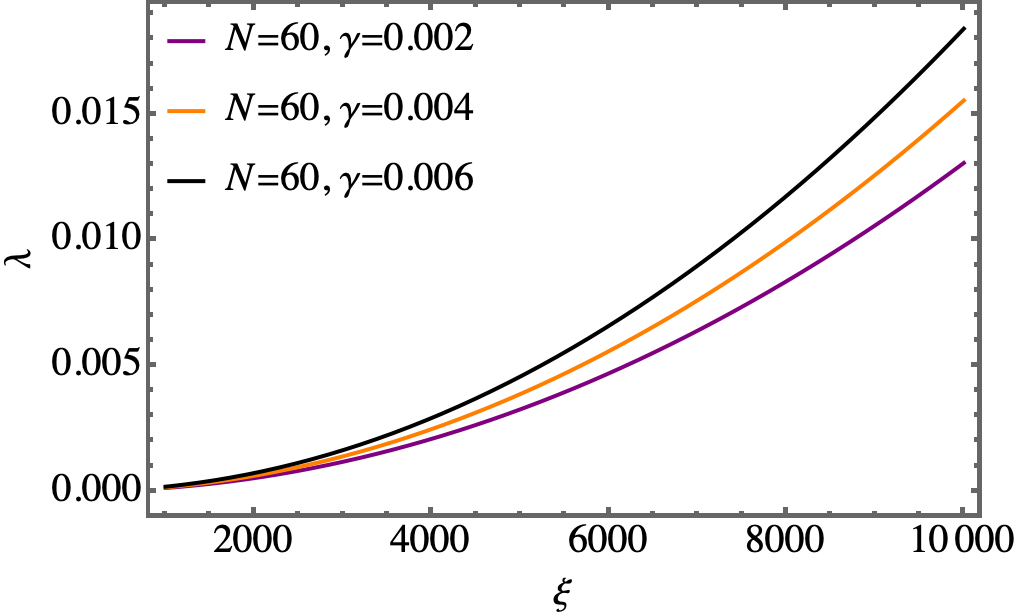}
\includegraphics[width=8 cm]{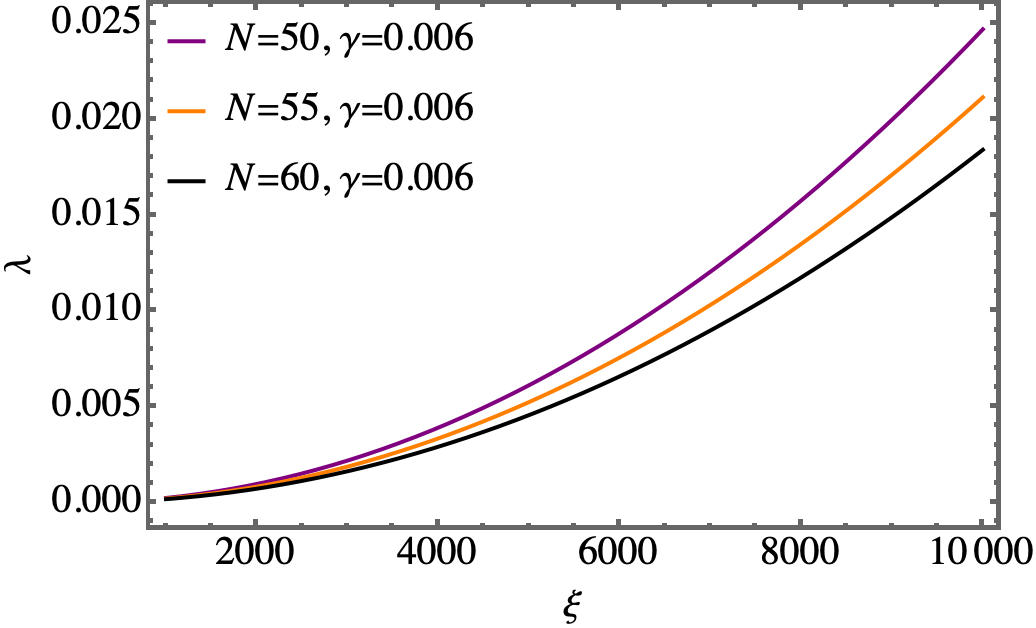}
\caption{Here we show (\ref{lamdxi}) as a function of $\xi$ for different values of the quantum correction parameter $\gamma$, at a fixed number of e-folds $N=60$ (upper panel) and for a fixed quantum correction $\gamma=0.006$, while varying the number of e-folds $N$ (lower panel).}\label{lambdaxi}
\end{figure}

%%%%%%%%%%%%%%%%%%%
We are now ready to compare the inflationary potential with experimental data. As a first step, we consider the constraints imposed by the measured amplitude of density perturbations, $A_{s}$~\cite{Planck:2013jfk}. To reproduce the correct value of $A_{s}$, the potential must satisfy the condition at horizon crossing, $\varphi_{*}$:
\begin{eqnarray}
A_{s}=\frac{1}{24\pi^{2} M^{4}_{p}}\Bigg|\frac{U_{*}}{\varepsilon_{*}}\Bigg|=2.2
\times 10^{-9}\,,
\end{eqnarray}
which implies
\begin{eqnarray}
\Bigg|\frac{U_{*}}{\varepsilon_{*}}\Bigg|=(A\,M_{p})^{4}=(0.0269\,M_{p})^{4}\,.\label{Uep}
\end{eqnarray}
In the case of a minimally coupled quartic potential, this requirement places a stringent condition on the self-coupling, which must take an unnaturally small value of $\lambda \sim 10^{-13}$~\cite{Guth:1982ec}. However, in the
present case, the above expression yields a relation between $\xi,\,\lambda$ and $\gamma$. We obtain from Eq.(\ref{Uep}):
\begin{eqnarray}
\lambda=\frac{4 A \gamma ^2 \xi ^2 \left(\gamma +e^{\frac{4 \gamma  M}{3}}\right)^2 }{3 \left(e^{\frac{4 \gamma  M}{3}}-1\right)^2}\left(\frac{\sqrt{\frac{e^{\frac{4 \gamma  M}{3}}-1}{\gamma }}}{\sqrt{\xi }}\right)^{-4 \gamma }\,.\label{lamdxi}
\end{eqnarray}
The resulting constraint is plotted in Fig.\ref{lambdaxi}. The Fig.\ref{lambdaxi} shows the relationship between the non-minimal coupling parameter $\xi$ (horizontal axis) and the self-coupling $\lambda$ (vertical axis) for different values of the quantum correction parameter $\gamma$, at a fixed number of e-folds $N=60$. The figure illustrates the interplay between non-minimal coupling and quantum corrections in determining viable inflationary scenarios. Larger $\xi$ values relax the smallness of $\lambda$, while higher $\gamma$ strengthens this trend. Thus, the plot provides evidence that quantum corrections allow inflation to be realized at more natural parameter values than in the purely classical $\varphi^{4}$ scenario. 
We also display the dependence of the self-coupling $\lambda$ on the non-minimal coupling parameter $\xi$ for a fixed quantum correction $\gamma=0.006$, while varying the number of e-folds $N$. It shows that both $\xi$ and $N$ critically determine the allowed values of  $\lambda$, providing guidance when matching theoretical models to observational constraints.

Next we consider the scalar spectral index $n_s$ and the tensor-to-scalar power ratio $r$. We have
\begin{eqnarray}
r\equiv16\varepsilon_{*}&\simeq& 16\bigg(\frac{8 \gamma ^2}{\kappa ^2}+\frac{8 M_{p}^4}{\kappa ^2 \left(\xi  \phi ^2\right)^2}+\frac{16 \gamma M_{p}^2}{\kappa ^2 \left(\xi  \phi ^2\right)}\bigg)\nonumber\\&=&\frac{12}{N^2}+\frac{16 \gamma }{N}+\frac{80 \gamma ^2}{9}+{\cal O}(\gamma^3)\,.\label{r}
\end{eqnarray}
and
\begin{eqnarray}
n_{s}&\equiv& 1-6\varepsilon_{*}+2\eta =16\varepsilon_{*}\nonumber\\&\simeq& 1-\frac{16 M_{p}^4+16 M_{p}^2 \xi  \phi ^2}{\kappa ^2 \xi ^2 \phi ^4}-\frac{32 \gamma  M_{p}^2}{\kappa ^2 \xi  \phi ^2}-\frac{16 \gamma ^2}{\kappa ^2}\nonumber\\&=&1-\frac{2}{N}-\frac{1.5}{N^2}+\bigg(1.33-\frac{2}{N}\bigg)\gamma\nonumber\\&&-0.0740741\bigg(15+4N\bigg)\gamma^{2}+{\cal O}(\gamma^3)\,.\label{ns}
\end{eqnarray}
By combining baryon acoustic oscillation (BAO) data~\cite{eBOSS:2020yzd} with 
CMB lensing measurements~\cite{Planck:2018lbu}, Ref.~\cite{Tristram:2021tvh} 
reported an improved constraint on the tensor-to-scalar ratio, 
$r < 0.032$ (95\% C.L.), compared to the slightly weaker bound 
$r < 0.038$ (95\% C.L.) obtained by P-ACT-LB-BK18~\cite{ACT:2025tim}. 
Using Eq.~(\ref{r}), this translates into an upper limit for $\gamma$:  
\begin{eqnarray}
\gamma < 0.06 \sqrt{\frac{N^2-150}{N^2}}-\frac{0.9}{N} \,,
\end{eqnarray}
which, for $N=60$, yields $\gamma < 0.044$. From Eq.~(\ref{ns}), the spectral index value $n_s = 0.9743$ can be reproduced for  
\begin{eqnarray}
\gamma \rightarrow 0.00674134, \quad \gamma \rightarrow 0.0624955 \,,
\end{eqnarray}
with the latter solution being phenomenologically disfavored. 
The addition of P-ACT data slightly shifts the preferred value of $n_s$ upward, 
as shown by the green contour. For $\gamma = 0$, the predictions coincide with those 
of the Starobinsky $R^2$ model and Higgs or Higgs-like inflation. However, in the 
range $50 < N < 60$, these models exhibit a tension with the P-ACT-LB measurement 
of $n_s$, at a level of approximately $\gtrsim 2\sigma$.
\begin{figure}
\includegraphics[width=8 cm]{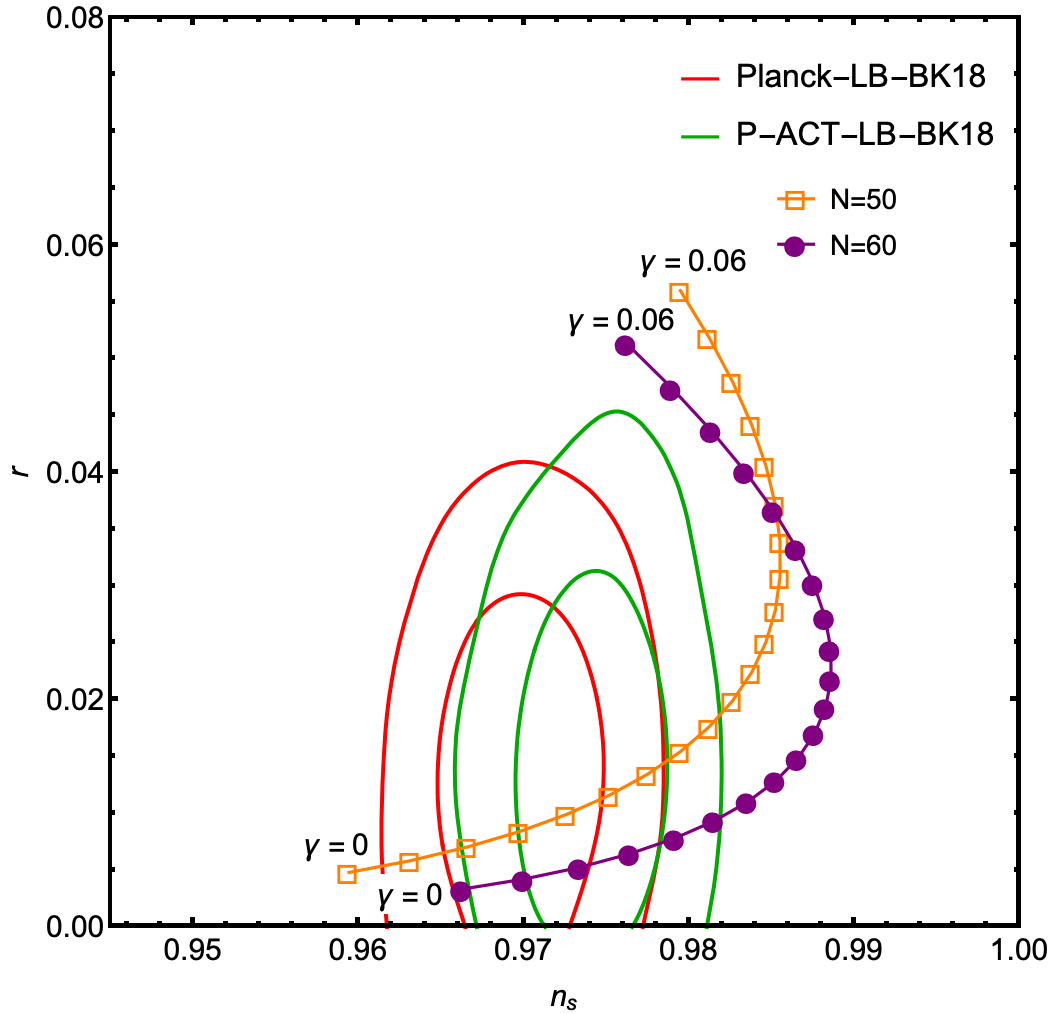}
\includegraphics[width=8 cm]{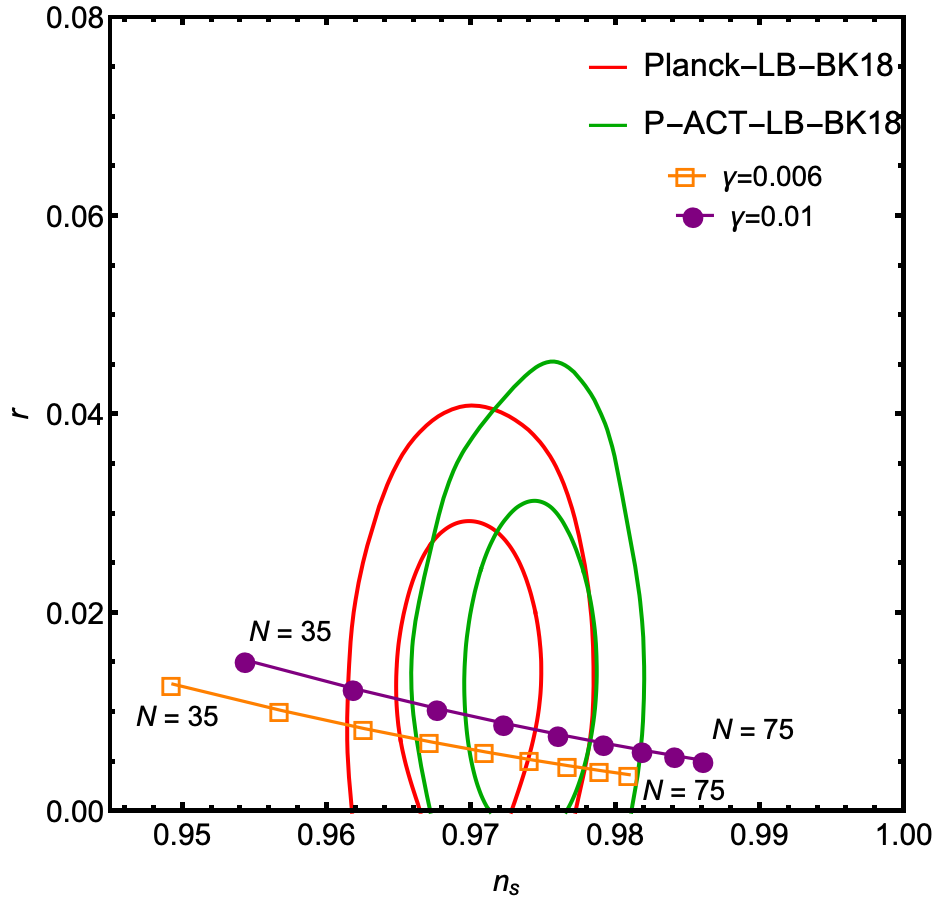}
\caption{Predictions for the present case, given for diﬀerent values
of $\gamma$ and $N$. The standard $\varphi^{4}$-Inflation is obtained for $\gamma=0$. We show the predictions for different values of the quantum correction parameter $\gamma$, at a fixed number of e-folds $N=50,\,60$ (upper panel), and for a fixed quantum correction $\gamma=0.006,\,0.01$, while varying the number of e-folds $N$ (lower panel).}\label{rns}
\end{figure}

The Fig.(\ref{rns}) highlights the impact of both the quantum correction parameter 
$\gamma$ and the number of e-folds $N$ on the inflationary predictions 
in the $(n_s,r)$ plane. For $\gamma=0$, the model reduces to predictions 
consistent with the Starobinsky $R^2$ scenario and Higgs(-like) inflation, 
yielding small tensor-to-scalar ratios and spectral indices aligned with 
Planck constraints. As $\gamma$ increases, the predictions shift toward 
higher values of $n_s$ and $r$, tracing upward trajectories. This trend 
becomes more compatible with the P-ACT-LB-BK18 contours, which favor 
slightly larger $n_s$ values than those preferred by Planck. Models with $N=50$ generate larger 
tensor-to-scalar ratios, moving closer to the observational upper bounds, 
while $N=60$ predictions fall within safer regions of parameter space, 
providing a better fit to the combined datasets. Overall, the results 
demonstrate that a modestly non-zero $\gamma$ broadens the phenomenological 
viability of the scenario, allowing it to accommodate both Planck and 
P-ACT data, with longer inflationary durations ($N \sim 60$) being particularly favored.

%%%%%%%%%%%%%%%%%%%
\section{Conclusions}\label{Con}
%%%%%%%%%%%%%%%%%%
In this work, we have revisited the inflationary dynamics of marginally deformed Starobinsky gravity in light of the latest observational constraints, particularly those arising from the ACT, DESI, and Planck collaborations. By incorporating quantum-induced deformations of the $R^{2}$ term, parametrized through a small correction $\gamma$, we analyzed the resulting scalar spectral index $n_{s}$ and tensor-to-scalar ratio $r$ within the standard slow-roll framework.

Our results show that even modestly nonzero values of $\gamma$ shift the predictions of the Starobinsky $R^{2}$ model toward higher $n_{s}$ and $r$, thereby easing the tension with the ACT+DESI (P-ACT-LB) constraints that report $n_{s} \simeq 0.9743 \pm 0.0034$. Importantly, we found that for $N \simeq 60$ $e$-foldings, the model accommodates both the Planck and ACT datasets, while shorter inflationary durations ($N \simeq 50$) yield larger tensor amplitudes, placing the scenario closer to the upper observational bounds. The analysis also highlights that quantum corrections relax the extreme fine-tuning of the quartic self-coupling $\lambda$ required in minimally coupled models, enabling more natural parameter choices when linked to the non-minimal coupling $\xi$.

Furthermore, the confrontation with current observational limits indicates that the parameter space with $\gamma \sim \mathcal{O}(10^{-3})$–$\mathcal{O}(10^{-2})$ remains viable, broadening the phenomenological applicability of Starobinsky-like inflation. For $\gamma = 0$, the framework reduces to the original $R^{2}$ scenario, which is in tension with ACT results at the $\gtrsim 2\sigma$ level, emphasizing the importance of marginal deformations in maintaining consistency with evolving data.

Overall, our study demonstrates that quantum-deformed extensions of the Starobinsky model provide a simple yet robust mechanism to reconcile inflationary predictions with the latest cosmological observations. Future CMB surveys, such as the Simons Observatory and CMB-S4, along with upcoming gravitational-wave experiments, will play a decisive role in testing these predictions and constraining the deformation parameter $\gamma$ with unprecedented precision.

\end{document}